\RequirePackage{fix-cm}
\documentclass[twocolumn,epjc3]{svjour3}  
\smartqed  
\RequirePackage{graphicx}
\RequirePackage[numbers,sort&compress]{natbib}
\RequirePackage[colorlinks,citecolor=blue,urlcolor=blue,linkcolor=blue]{hyperref}
\usepackage{amsmath}
\RequirePackage{graphicx}
\journalname{Eur. Phys. J. C}
\begin{document}

\title{Measuring Cosmic Growth Rate with CSST Spectroscopic Survey and Fast Radio Burst
}


\author{Shi-Yuan Wang\thanksref{addr1,addr2}
        \and
        Jun-Qing Xia\thanksref{e2,addr1,addr2} 
}

\thankstext{e2}{e-mail: xiajq@bnu.edu.cn}


\institute{Institute for Frontiers in Astronomy and Astrophysics, Beijing Normal University, Beijing 100875, China \label{addr1}
           \and
           School of Physics and Astronomy, Beijing Normal University, Beijing 100875, China \label{addr2}
}

\date{Received: date / Accepted: date}

\maketitle

\begin{abstract}
The cosmic growth rate, which is related to peculiar velocity and is a primary scientific objective of galaxy spectroscopic surveys, can be inferred from the Redshift Space Distortion effect and the kinetic Sunyaev-Zel’dovich (kSZ) effect. However, the reconstruction noise power spectrum of the radial velocity field in kSZ is significantly dependent on the measurement of the small-scale galaxy-electron power spectrum $P_\text{ge}$. In this study, we thoroughly discuss the enhancement of cosmic growth rate measurements facilitated by Fast Radio Bursts (FRBs), which probe the electron density of the universe along their propagation paths to provide crucial additional information on $P_\text{ge}$. Subsequently, we utilize future spectroscopic surveys from the Chinese Space Station Telescope and the CMB-S4 experiment, combined with FRB dispersion measures, to achieve precise measurements of the cosmic growth rate at redshifts $z_\text{g} = 0.15,\,0.45,\,0.75$. Employing Fisher matrix forecasting analysis, we anticipate that constraints on $f\sigma_8$ will reach a precision of 0.1\% with a sample size of $10^6$ FRBs. Furthermore, we perform a global analysis using Markov Chain Monte Carlo methods to constrain key parameters of three distinct dark energy models and a modified gravity model based on cosmic growth rate measurements. The results demonstrate that these refined $f\sigma_8$ measurements considerably enhance the constraints on relevant cosmological parameters compared to those obtained from Planck CMB data. As the number of observed FRBs increases, alongside more precise galaxy surveys and next-generation CMB observations, new opportunities will arise for constraining cosmological models using the kSZ effect and for developing novel cosmological applications of FRBs.
\end{abstract}

\section{Introduction} \label{sec:intro}

The growth of large-scale structures are critically influenced by the underlying gravitational theory, as the growth function varies across cosmological models depending on the gravitational frameworks \cite{2023A&ARv..31....2H}. As a result, precise measurements of this function provide a robust test for the standard cosmological model and its alternatives \cite{2020JCAP...10..042L, 2020EPJC...80.1210V,2023A&ARv..31....2H}. The growth rate is commonly expressed as $f = d\ln D / d\ln a$, where $D$ denotes the linear growth function of the matter density contrast. Observationally, the growth rate is often determined in terms of the combination $f\sigma_8$, where $\sigma_8$ represents the amplitude of matter density fluctuations on a scale of 8 Mpc/h \cite{2017PhRvD..96b3542N}. This combination is particularly valuable due to its independence from galaxy bias, rendering it directly observable through observational techniques such as weak gravitational lensing or redshift space distortion (RSD).

Additionally, the peculiar velocity is highly sensitive to structure growth \cite{2014MNRAS.445.4267K}. In linear perturbation theory, the peculiar velocity field is directly related to the cosmic growth rate and the galaxy density field \cite{1980lssu.book.....P}. This relationship is expressed as $\mathbf{v} = {i \mathbf{k} a f H \delta} /{k^{2}}$, indicating that observations of the peculiar velocity field provide an effective indirect probe of the growth rate of cosmic structures.

Currently, one of the widely utilized methods for measuring the growth rate is through the kinetic Sunyaev-Zel’dovich (kSZ) effect, first observed by \cite{2012PhRvL.109d1101H}. The kSZ effect is a cosmological phenomenon related to the Cosmic Microwave Background (CMB) radiation. When galaxy clusters move relative to the CMB, meaning they possess a non-zero peculiar velocity, the CMB photons interacting with the free electrons within these clusters undergo a frequency shift—either a blue shift or a redshift—depending on the direction of the cluster's motion relative to the observer. The kSZ effect does not change the total number of CMB photons, and it preserves the blackbody spectrum of the CMB. By combining observations of the kSZ effect with data from galaxy surveys such as DES \cite{2023PhRvD.108b3516M}, BOSS \cite{2018PhRvD..97b3514L}, SDSS \cite{2021A&A...645A.112T} and the Roman Space Telescope \cite{2023PhRvD.107h3502F}, we can study the motion of large-scale structures in the universe, including the velocity field of galaxy clusters, mean optical depth and the formation and evolution of cosmic large-scale structures. This approach provides valuable insights into the distribution of dark matter and the history of cosmic expansion \citep{alonso2016reconstructing, farren2022ultralight}.

\cite{2018arXiv181013423S} demonstrated that by combining galaxy surveys with CMB data, one can reconstruct the peculiar velocity field and its noise power spectrum using a quadratic estimator. This estimator depends on the electron-galaxy velocity power spectrum $P_\text{ge}(k_S)$ and the galaxy auto-power spectrum $P_\text{gg}(k_S)$. However, the observed power spectrum in kSZ measurements is actually a bispectrum, representing the product of $P_\text{ge}(k_S)$ and $P_\text{gv}(k_L)$, rather than a direct measurement of these two components. Therefore,  it is impossible to directly measure the small-scale electron-galaxy power spectrum $P_\text{ge}(k_S)$ independently.
In practical experiments, a fiducial model $P_\text{ge}^{\text{fid}}(k_S)$ is introduced, which may differ from the true power spectrum $P_\text{ge}^{\text{true}}(k_S)$. This discrepancy leads to a proportional difference between the reconstructed and true velocities, requiring the incorporation of a large-scale linear bias $b_\text{v}$ in the relationship between the peculiar velocity field and the growth rate \cite{2024arXiv240500809B}. This bias originates from the uncertainty in $P_\text{ge}(k_S)$ in kSZ measurements.

Fast Radio Bursts (FRBs) are extragalactic radio transient phenomena characterized by significant signal dispersion that correlates with their distance. These signals undergo significant dispersion due to the ionized medium present along their propagation path. The observed dispersion, quantified by the Dispersion Measure (DM), is significantly exceeds the contribution from the local environment \cite{2023RvMP...95c5005Z}. It is well established that the vast majority of FRBs originate from extragalactic sources, making them valuable cosmic probes. Owing to this characteristic, FRBs can be utilized to detect the distribution of free electrons along their signal path. Consequently, FRBs can be employed to study the reionization history \cite{2024arXiv240509506P, 2022ApJ...933...57H, 2021JCAP...05..050D}, locate missing baryons \cite{2021MNRAS.503.4576D, 2020Natur.581..391M}, dark matter \cite{2023ApJ...950...53H, 2020ApJ...900..122S}, fine-structure constant evolution \cite{2024arXiv240611691L} or other fundamental physics\cite{2023APS..APRS04004Y}. By correlating the DM field with the galaxy field, an additional measurement of the electron-galaxy cross-correlation function can be achieved with a high signal-to-noise ratio \cite{2023arXiv231101899N,2019PhRvD.100j3532M}. With precise measurements of $P_\text{ge}(k_S)$ on small scales, it would be possible to obtain an accurate velocity bias $b_\text{v}$, thus improving the precision of growth rate measurements derived from the kSZ effect.

In this study, we investigate how cosmic growth rate measurements can be improved using additional information from $P_\text{ge}$ from FRBs. We utilize forthcoming spectroscopic surveys from the Chinese Space Station Telescope (CSST) and the CMB-S4 experiment, combined with FRB dispersion measures, to achieve precise measurements of cosmic growth rate. We then perform a global analysis using Markov Chain Monte Carlo (MCMC) methods to constrain key parameters of three distinct dark energy models: the $\Lambda$CDM model, the constant equation of state dark energy model, and the dynamical dark energy model, informed by cosmic growth rate measurements.

Specifically, in Sec.\ref{sec:method}, we review conventional approaches to measuring the growth rate via the kSZ effect. In Sec.\ref{sec:power}, we calculate the power spectrum related to FRB DM, along with the associated signal-to-noise ratio (SNR) and uncertainties, leading to a refined estimate of $b_\text{v}$. In Sec.\ref{sec:growth}, we present the results of constraints on the growth rate obtained from kSZ and RSD measurements. Finally, in Sec.\ref{sec:constraint}, we outline the MCMC baseline and utilize the previously obtained $f\sigma_8$ results to constrain different cosmological models and the modified gravity model. In this work, we assume the standard $\Lambda$CDM cosmological model \cite{2020A&A...641A...6P} with the following parameters: $\Omega_\text{b}h^2 = 0.022$, $\Omega_{\rm c}h^2 = 0.120$, $100\theta_{\rm MC} = 1.041$, $\tau = 0.054$, $\ln(10^{10}A_{\rm s}) = 3.0449$, and $n_{\rm s} = 0.966$.

\section{Measurements of growth rate with kSZ effect}\label{sec:method}

\subsection{CSST spectroscopic survey}

The CSST is a two-meter space telescope equipped with three spectroscopic bands—GU, GV, and GI—covering a wavelength range from approximately 250 to 1000 nm. The CSST is designed to survey a 17,500 deg$^2$ area, capturing spectral information for over one hundred million galaxies. The redshift distribution of the CSST galaxy spectroscopic survey, which is simulated based on the zCOSMOS catalog \citep{2007ApJS..172...70L,2009ApJS..184..218L}, spans a range from 0 to 2.5, with a peak at $z_\text{g} = 0.3 - 0.4$. For practical analysis, the galaxy sample is typically divided into five tomographic redshift bins with an interval of $\Delta z_\text{g} = 0.3$, assuming no spectroscopic redshift outliers for simplicity. In this study, we focus on the first three redshift bins, specifically $z_\text{g}$: [0,0.3], [0.3,0.6], and [0.6,0.9], with corresponding galaxy number densities $n_\text{g}^{\text{ori}} = 3.4 \times 10^{-2}$, $1.1 \times 10^{-2}$, and $5.5 \times 10^{-3} \text{(h/Mpc)}^3$, respectively. Additionally, we adopt a galaxy bias model \(b_\text{g} = 1+0.84z_\text{g}\), and the sky coverage of survey \(f_{\text{sky}}\simeq 0.424\) \citep{2019ApJ...883..203G}.

\subsection{kSZ Effect power spectrum}

The kSZ effect is an effective probe of the radial velocity field of matter \citep{2024arXiv240500809B}. Specifically, the power spectrum of the peculiar velocity field can be utilized to infer the growth rate of cosmic structures \citep{2019PhRvD.100j3532M}:
\begin{equation}
	P^{\rm kSZ}_{\rm{gg}}  =(b_\text{g}+f(z)\mu^2)^2P_{\text{mm}}(k,z)
\end{equation}
\begin{equation}\label{eq:pgv}
	P^{\rm kSZ}_{\rm{gv}} =b_\text{v}\left(\frac{f(z)aH(z)}k\right)(b_\text{g}+f(z)\mu^2)P_{\text{mm}}(k,z)
\end{equation}
\begin{equation}\label{eq:pvv}
	P^{\rm kSZ}_{\rm{vv}} =b_\text{v}^{2}\left(\frac{f(z) a H(z)}{k}\right)^{2} P_{\text{mm}}(k,z)
\end{equation}
where the $b_\text{g}$ we employ a CSST-like bias mentioned above.

Furthermore, as shown by \cite{2018arXiv181013423S}, the reconstruction noise power spectrum of the radial velocity field in the kSZ effect can be expressed as:
\begin{equation}\label{eq:nvv}
    N_\text{vv} = \frac{\chi_\text{g}^2}{\mu^{2}K_\text{g}^2} \left[\int \frac{k_S \, dk_S}{2\pi} \left(\frac{P_{\text{ge}}(k_S)^2}{P_{\text{gg}}^{\text{tot}}(k_S) \, C_{\ell=k_S \chi_\text{g}}^{\text{TT,tot}}}\right)\right]^{-1}
\end{equation}
where \( P_{\text{ge}}(k_S) \) is the small-scale galaxy-electron cross-power spectrum, and \( P_{\text{gg}}^{\text{tot}}(k_S) \) is the small-scale galaxy auto-power spectrum, including the noise term. We consider the small-scale range \( k_S \sim (0.1, 10) \text{Mpc}^{-1} \). Additionally, \( C_{\ell}^{\text{TT,tot}} \) denotes the total CMB power spectrum, encompassing the intrinsic CMB power spectrum with the experimental noise.

In our calculations, we primarily focus on the CSST spectroscopic survey and the CMB-S4 survey \citep{2016arXiv161002743A}. We model the CMB noise spectrum as follows:
\begin{equation}\label{eq:nl}
    N_{\ell} = \Delta_{T}^{2} \exp \left[\frac{\ell(\ell+1) \theta_{\mathrm{FWHM}}^{2}}{8 \ln 2}\right]
\end{equation}
Here, we use a beam size of \( \theta_{\text{FWHM}} = 1.5 \) arcmin and an effective white noise level of \( 1.8 \, \mu\text{K} \) arcmin. This model does not account for foreground cleaning.

\begin{figure}[t]
    \centering
    \includegraphics[width=3in]{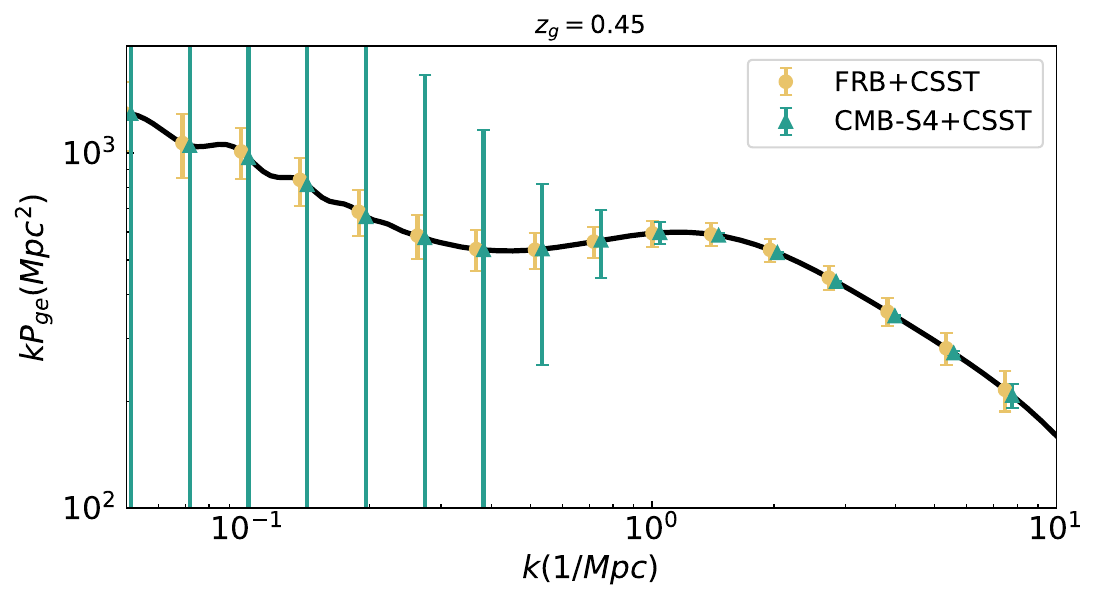}
    \caption{ Galaxy-electron power spectra and their uncertainties in \( k \)-bins, comparing the FRB+CSST (yellow) and CMB-S4+CSST (green) approaches. }\label{fig:kpge}
\end{figure}

\subsection{The uncertainty of $P_\text{ge}$}

To calculate the power spectrum \( P_\text{ge}(k) \), we will focus on estimating its uncertainty across different \( k \)-bins. Clearly, the kSZ effect can provide information about the electron distribution through its influence on the CMB \citep{2018arXiv181013423S}:
\begin{equation}\label{eq:dpge1}
	\Delta P^{\rm kSZ}_{\text{ge}}=\left[V_\text{g} \frac{K_{g}^{2}}{12 \pi^{3} \chi_\text{g}^{2}}\left(\int d k_{L} k_{L}^{2} \frac{P_\text{gv}\left(k_{L}\right)^{2}}{P_\text{gg}^{\text {tot}}(k_{L})}\right)\right.
\end{equation}
\begin{displaymath}
	\left.\times\left(\int_{k_{S}^{\min }}^{k_{S}^{\max }} d k_{S}  \frac{k_{S}}{P_\text{gg}^{\text {tot}}(k_{S})} \frac{1}{(C_{\ell}^{\text {TT,tot}})_{\ell=k_{S} \chi_\text{g}}}\right)\right]^{-1 / 2}
\end{displaymath}
where \( V_\text{g} = \chi_\text{g}^3 \) represents the comoving volume, while \( K(z) \equiv -T_{\mathrm{CMB}} \sigma_{T} n_\text{e0} x_\text{e} e^{-\tau(z)}(1+z)^{2} \) is the kSZ radial weight function at redshift \( z \). Here, \( \sigma_\text{T} \) is the Thomson cross section, \( n_\text{e0} \) denotes the present-day comoving electron density, \( x_\text{e} \) is the ionization fraction, and \( \tau(z) \) is the optical depth. The total galaxy power spectrum \( P_\text{gg}^{\text{tot}}(k) \) consists of the intrinsic galaxy power spectrum along with the shot noise contribution. Additionally, the total CMB temperature power spectrum \( C_\ell^{\text{TT,tot}} \) is configured as specified in Eq.(\ref{eq:nl}).

On the one hand, the results for the power spectrum \(P_\text{ge}\) and its uncertainties across different \(k\)-bins at \(z_\text{g}=0.45\), along with the kSZ measurement represented by the green line, are illustrated in Figure \ref{fig:kpge}. { For FRBs catalog, we set the parameters with \(\sigma_\text{DM} = 300 \, \text{pc}/\text{cm}^3\) and \( N_\text{FRB} = 10^4 \).} It is clear that only kSZ measurement is limited to measuring \(P_\text{ge}\) effectively within a narrow range of scales in the 1-halo regime. Moreover, the traditional kSZ-based approach assumes prior knowledge of \(P_\text{gv}\), which may not always be available, leading to potential inaccuracies.

On the other hand, from Eqs.(\ref{eq:pgv}-\ref{eq:nvv}), it is evident that the growth rate derived using the kSZ method is significantly affected by the velocity bias \( b_\text{v} \). This bias originates from the discrepancy between the fiducial \( P_\text{ge}^\text{fid} \) and the true \( P_\text{ge}^\text{true} \) in kSZ measurements, and is influenced by the integration of \( P_{\text{ge}} \) on small scales. 

As a result, the kSZ method alone may not provide an accurate determination of \( b_\text{v} \). To improve the accuracy of growth rate measurements, it is crucial to supplement the kSZ data with additional information about \( P_{\text{ge}} \) and \( b_\text{v} \). With this extra information, the precision of the growth rate estimates can be significantly enhanced.

\section{Fast Radio Bursts}\label{sec:power}


\subsection{Redshift Distribution}

Since the DM is the result of integrating the electron density along the line of sight, it effectively traces the fluctuations in the electron distribution. As FRBs travel through galaxies, the free electrons within these galaxies contribute to the observed DM, making the cross-correlation of DM with galaxies a powerful tool for probing the free electron content of these galaxies. For this approach to be effective, the galaxies should be in the foreground, with the FRBs located behind them. This setup allows the DM to trace the free electron distribution in the galaxies, which is the primary focus of our analysis. To achieve this, we assume that the FRBs are positioned \( \Delta z_\text{f} = 0.1 \) behind each galaxy redshift bin. This assumption ensures that the FRBs lie sufficiently behind the galaxies, enabling a meaningful cross-correlation that can reveal the distribution of free electrons within the galaxies \citep{2019PhRvD.100j3532M, 2018arXiv181013423S}.

To analyze the correlation between galaxies and DM of FRBs, we consider two-dimensional DM fields where FRBs are located in three distinct background shells with central redshifts \(z_\text{f} = 0.4\), \(z_\text{f} = 0.7\), and \(z_\text{f} = 1.0\). These shells are positioned behind three galaxy bins, each corresponding to thin redshift shells with mean redshifts \(z_\text{g} = 0.15\), \(z_\text{g} = 0.45\), and \(z_\text{g} = 0.75\). The redshift width of each galaxy shell is \(\Delta z_\text{g} = 0.3\).

This setup ensures that there is a sufficient separation between the foreground galaxy shells and the background FRB shells within each group. As a result, the correlations observed between galaxies and DM are purely due to the free electrons within the galaxies along the line of sight, without interference from any spatial correlations between the galaxies and the host galaxies of the FRBs. 

\subsection{Angular Power Spectrum}

Given that FRBs originate from cosmological distances, the DM can be expressed as \citep{2023RvMP...95c5005Z}:
\begin{equation}
\text{DM} = \int_{0}^{D_{z}} \frac{n_e(l)}{1+z(l)} \, dl
\end{equation}
where \( n_e(l) \) is the local electron number density along the line of sight, \( dl \) is the physical distance element, which can be expressed as: $ dl =  {c}/{(1+z)/H(z)} \, dz$. Here \( H(z) \) is the Hubble parameter at redshift \( z \), \( D_z \) is the physical distance corresponding to redshift \( z \).

If we assume that galaxies lie in thin redshift shells ($\Delta z_\text{g}=0.3$) centered at $z_\text{g}$, we can use the Limber approximation \citep{1953ApJ...117..134L} to obtain the cross angular power spectra of galaxy and DM:
\begin{equation}
	C_{\ell}^{\rm Dg} =  n_\text{e0}\frac{(1+z)}{\chi_\text{g}^{2}} P_\text{ge}\left(k, z_{g}\right)_{k = \ell / \chi_\text{g}}
	\label{eq:cldg}
\end{equation}
where $P_\text{ge}$ 
can be calculated using halo models as described in \cite{2018arXiv181013423S}. The electron profile $u_e(k|m,z)$ 
is given by the “AGN” model from \cite{2016JCAP...08..058B}. This model accounts for the distribution of free electrons around galaxies, influenced by feedback from Active Galactic Nuclei (AGN). $n_\text{e0}$ denotes the mean number density of free electrons today, $\chi_\text{g}$ represents the assumed comoving side length of the observed box universe within each galaxy redshift bin. { Notably, the electron profile derived in this model inherently carries some uncertainty. However, on the scales of interest in this study, this uncertainty is negligible compared to \(\sigma_\text{DM}\). Future high-resolution FRB observations may require a more careful consideration of this error.}

When observing the two-dimensional DM field using a discretely sampled catalog of FRBs, the observations are subject to noise. This noise can be quantified using a noise power spectrum, which accounts for the discrete sampling and the variance in the observed DMs. The noise power spectrum, $N_{\rm DD}$, can be expressed as: $N_{\rm DD} = \sigma_\text{DM}^{2} / n_\text{f}^\text{2d}$, where $n_\text{f}^\text{2d}$ is the number density of FRBs per steradian. $\sigma_\text{DM}^2$ is the total variance of the DMs, which includes contributions from both intrinsic variations in the DMs of FRBs and observational uncertainties. 
In our study, we consider three different cases for the total variance of the DMs: \(\sigma_{\rm DM} = 100, 300, 1000 \, \text{pc}/\text{cm}^3\). These cases are used to assess the influence of noise on the statistical analysis, helping to understand how varying levels of DM variance affect the results.

\subsection{The uncertainty of $P_{\text{ge}}$}

Apart from kSZ effect, the second approach to calculate \(\Delta P_\text{ge}\) involves utilizing FRBs in conjunction with observed galaxies. By analyzing the cross-correlation between the FRB DM field and the galaxy distribution, we can derive \(\Delta P_\text{ge}\) \citep{2019PhRvD.100j3532M}:
\begin{equation}\label{eq:dpge2}
	\Delta P^{\rm FRB}_{\mathrm{ge}}=\frac{\chi_\text{g}}{n_\text{e0}(1+z_\text{g})}\left(\Omega \int_{\ell_{\min }}^{\ell_{\max }} \frac{l d l}{2 \pi} \frac{1}{\left({N}_{\ell}^\text{Dg}\right)^{2}}\right)_{\ell=k \chi_\text{g}}^{-1 / 2}
\end{equation}
where the noise term $({N}_{\ell}^{D g})^{2}$ is given by:
\begin{equation}\label{eq:nldg}
	({N}_{\ell}^\text{Dg})^{2} =(C_{\ell}^\text{gg}+N_\text{gg})(C_{\ell}^{\rm DD}+N_{\rm DD})
\end{equation}
where $C_{\ell}^{\rm gg}$ and $C_{\ell}^{\rm DD}$ are the auto angular power spectra of galaxies and DMs of FRB, $N_\text{gg}=1/n^{2d}_\text{g}$ represents the noise, defined as the inverse of the number density of CSST galaxies per steradian. In this analysis, we consider $10^4$ FRB DMs and the DM scatter of $\sigma^2_{\rm DM}=300{\rm pc/cm^3}$ for comparison.

In Figure \ref{fig:kpge}, we show \(P_\text{ge}\) and  its uncertainties calculated using FRB DM, represented by yellow lines. Compared to kSZ measurements, FRB DMs allow for accurate measurements of \(P_\text{ge}\) over a much broader range of scales. FRB DMs can break the degeneracy between \(P_\text{ge}\) and \(P_\text{gv}\), enabling a more precise estimation of \(P_\text{ge}\). This advantage arises because FRBs directly probe the electron density along their lines of sight, providing independent and supplementary information that reduces uncertainty in the power spectrum estimation. As a result, FRBs offer a promising approach to improving cosmic growth rate measurements, offering a new avenue for improving our understanding of cosmic structure formation and evolution.

\subsection{The uncertainty of $b_\text{v}$}

With more accurate \(P_\text{ge}\), it is meaningful to calculate $b_\text{v}$ using FRB DM, which arises from the discrepancy between \( P_\text{ge}^\text{true} \) and \( P_\text{ge}^\text{fid} \). To calculate the uncertainty of the velocity bias \(b_\text{v}\) for the kSZ effect, we can use the power spectrum \(P_\text{ge}\) obtained in the above section. In \cite{2019PhRvD.100j3532M}, the velocity bias is derived from integrating over the small-scale galaxy-electron power spectrum \(P_\text{ge}\), typically in the range \(k_S \sim (0.1, 10) \text{Mpc}^{-1}\). The integral for the velocity bias \(b_\text{v}\) can be expressed as:
\begin{equation}\label{eq:bv}
	b_\text{v}=\frac{\int d k_{S} F\left(k_{S}\right)P_{\mathrm{ge}}^{\mathrm{true}}\left(k_{S}\right)}{\int d k_{S} F\left(k_{S}\right) P_{\mathrm{ge}}^{\mathrm{fid}}\left(k_{S}\right)}~.
\end{equation}
The uncertainty in \(b_\text{v}\) is then derived from the uncertainty in \(P_\text{ge}\) over this range:
\begin{equation}
\sigma(b_\text{v})^2  = \frac{\int d k_S F\left(k_S\right)^2 G\left(k_S\right)^{-1}}{\left(\int d k_S F\left(k_S\right) P_{\mathrm{ge}}^{\mathrm{fid}}\left(k_S\right)\right)^2}
\end{equation}
where the terms
\begin{equation}
	G\left(k_S\right)=\left(\frac{\chi_\text{g}}{n_\text{e0}(1+z_\text{g})}\right)^{-2}\left(\frac{k_S \Omega}{2 \pi}\right)\left(\frac{1}{(N_{\ell=k_S \chi_\text{g}}^\text{Dg})^2}\right)
\end{equation}
and
\begin{equation}
	F\left(k_{S}\right)=k_{S} \frac{P_{\mathrm{ge}}^{\mathrm{fid}}\left(k_{S}\right)}{P_\text{gg}^{\mathrm{tot}}\left(k_{S}\right)}\left(\frac{1}{C_{\ell=k_S \chi_\text{g}}^{\mathrm{TT}, \mathrm{tot}}}\right)~.
\end{equation}
This approach allows us to estimate the uncertainty in \(b_\text{v}\) by taking into account the uncertainties in the power spectrum measurements and the integration over the relevant scales.

\begin{figure}[t]
    \centering
    \includegraphics[width=3in]{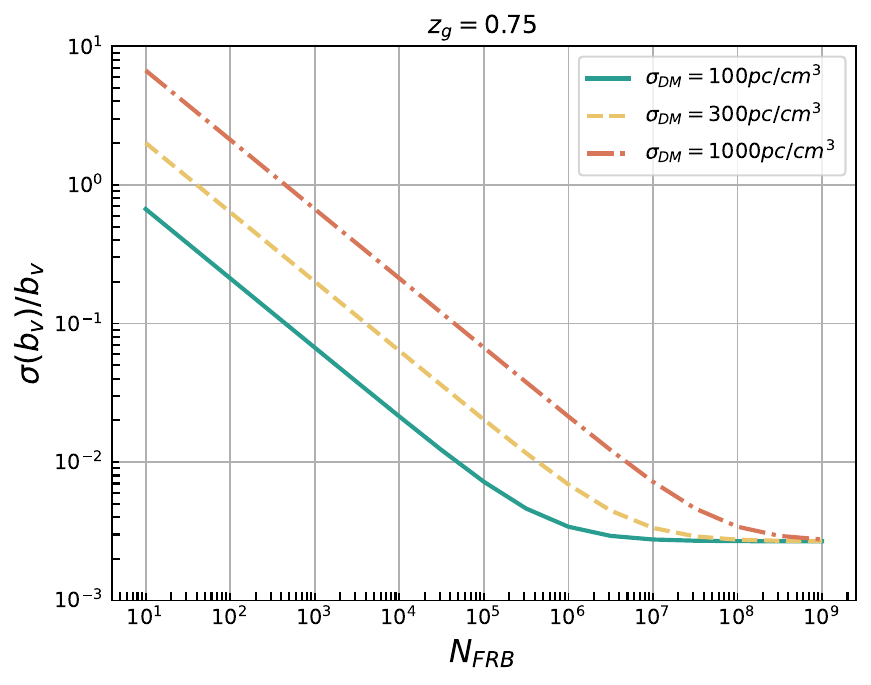}
    \caption{The uncertainty of kSZ velocity bias at redshift $z=0.75$, as functions of the number of FRBs. The solid, dashed, and dashed-dotted lines represent $\sigma_{\text{DM}}=1000,300,100\text{pc}/\text{cm}^3$, respectively. { The fiducial value is $b_{\rm v}=1$.} }
    \label{fig:bv75}
\end{figure}

In Figure \ref{fig:bv75}, the uncertainty on the velocity bias \(b_\text{v}\) is shown, with a fiducial value \(b_\text{v}=1\) to marginalize over \(b_\text{v}\). This figure illustrates the significant role that FRBs play in constraining \(b_\text{v}\) by providing additional information on the power spectrum \(P_\text{ge}\). The integration over a broad range of scales enabled by FRB data helps to reduce the uncertainty in \(b_\text{v}\), leading to more precise measurements compared to traditional methods that rely on limited scale ranges. As the number of FRBs increases, the uncertainty in the measurement decreases substantially, eventually stabilizing at approximately \(3 \times 10^{-3}\). This highlights the valuable contribution of FRBs in improving our understanding of cosmic growth rates and velocity biases.

\begin{figure*}[t]
    \centering
    \includegraphics[width=7in]{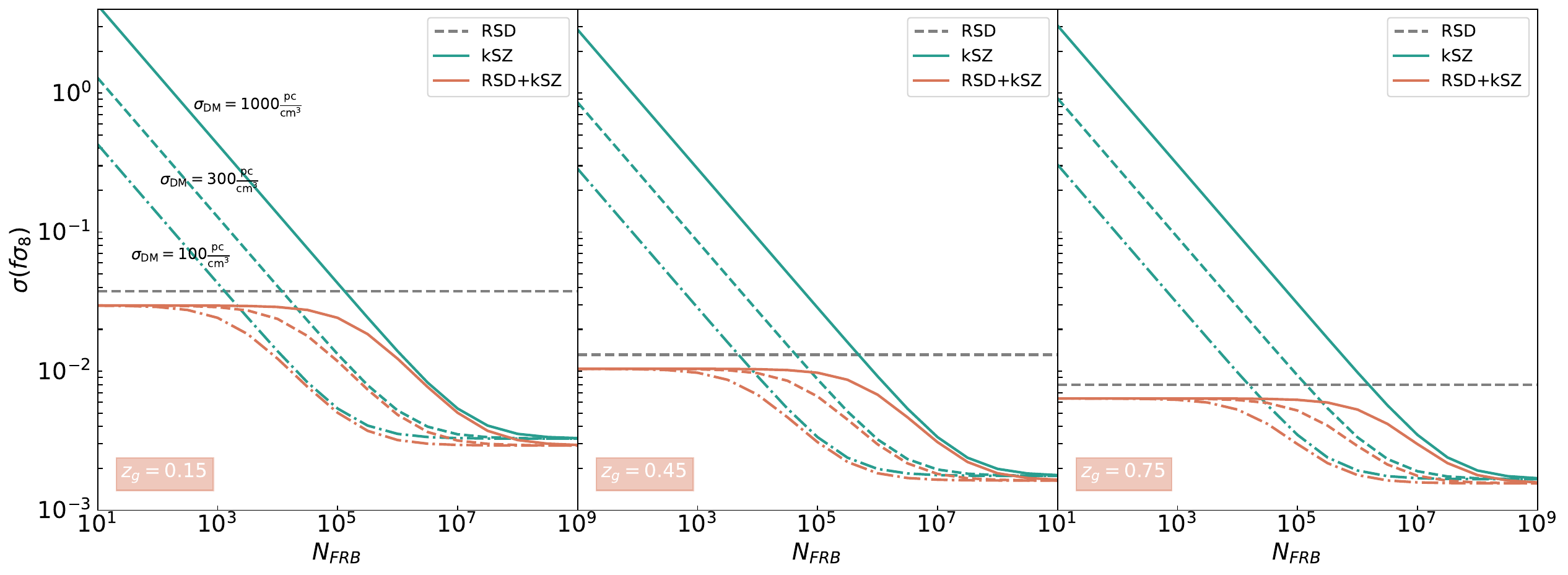}
    \caption{ The uncertainty in the combined measurement of \( f\sigma_8 \). The gray line represents the result from RSD. The green lines show the results from the kSZ effect with prior of \(b_\text{v}\), while the red lines represent the results from combining RSD with kSZ. The solid, dashed, and dash-dotted lines correspond to \(\sigma_{\text{DM}} = 1000, 300, 100 \, \text{pc}/\text{cm}^3\), respectively.}
    \label{fig:fs8}
\end{figure*}

\section{Results of the growth rate}\label{sec:growth}

Considering the experimental specifications of the CSST spectroscopic survey and the CMB-S4 survey, along with FRB DM measurements, we can forecast the cosmic growth rate and other cosmological parameters by constructing the Fisher matrix for the modes of the galaxy overdensity field and the reconstructed velocities. The information matrix for a redshift bin \(z_\text{g}\) can be expressed as:
\begin{equation}
	\mathcal{F}_{i j}=\frac{V_{\text {s}}}{8 \pi^{2}} \int_{-1}^{+1} d \mu \int_{k_{\min }}^{k_{\max }} k^{2} d k F_{i j}(k, \mu)
\end{equation}
where $V_s$ is the survey volume. 

For the kSZ measurement, in addition to the parameters \( f\sigma_8 \) and \( b_\text{g}\sigma_8 \), we also include the velocity bias \( b_\text{v} \). Given that FRB DMs provide additional information on \( b_\text{v} \), we incorporate the information about \( \sigma^2(b_\text{v}) \), which depends on the number of FRBs \( N_{\rm FRB} \) and the scatter \( \sigma_{\rm DM} \), into the information matrix. Including the galaxy overdensity field, the reconstructed velocities, and the priors on \( b_\text{v} \) provided by FRBs, the Fisher matrix for the kSZ measurement is given by:
\begin{equation}\label{eq:fisher_kSZ}
	\mathcal{F}_{ij}^\text{kSZ}=\frac{V_{\text {s}}}{8 \pi^{2}} \int_{-1}^{+1} d \mu \int_{k_{\min }}^{k_{\max }} k^{2} d k \text{Tr}\left[C_{, i} C^{-1} C_{, j} C^{-1}\right]  + \mathcal{F}^{\prime}
\end{equation}
where $\mathcal{F}^{\prime}$ is a $3\times$3 matrix in which only the third diagonal element is ${1}/{\sigma^2(b_\text{v})}$, while others are zeros, and $C$ is the covariance matrix which includes the sum of signal and noise power spectra \citep{2019PhRvD.100j3532M}:
\begin{equation}
    C=\left[\begin{array}{cc}
P_{\text{gg}}^\text{kSZ}+N_\text{gg} & P_{\text{gv}}^\text{kSZ} \\
P_{\text{gv}}^\text{kSZ} & P_{\text{vv}}^\text{kSZ}+N_\text{vv}~
\end{array}\right].
\end{equation}

In addition, we also consider the RSD measurement. Assuming a CSST-like galaxy survey with the same three redshift bins as kSZ measurement, the galaxy power spectrum is:
\begin{equation}
	P^{\rm RSD}_{\rm gg}(k,z,\mu) = P^{\rm RSD}_{\rm{gg,0}}(k,z,\mu) e^{-(k\mu\sigma_r)^2} + \frac{1}{\bar{n}_\text{g}} + N_{\text{sys}}
\end{equation}
where 
\begin{equation}
	P^{\rm RSD}_{\rm{gg,0}}(k, z, \mu)=\left[b_\text{g} \sigma_{8}(z)+f(z) \sigma_{8}(z) \mu^{2}\right]^{2} \frac{P_{\mathrm{mm}}(k, z=0)}{\sigma_{8}^2(z=0)}
\end{equation}
where $b_\text{g},f(z),\sigma_8(z)$ represent the galaxy bias, the cosmic growth rate, and the root mean square (RMS) fluctuations in spheres with a radius of 8 Mpc/h, respectively. There are two parameters involved: \( f\sigma_8 \) and \( b_\text{g}\sigma_8 \).

The information matrix \(\mathcal{F}_{ij}^\text{RSD}\) can be expressed in terms of the 3D galaxy power spectrum as follows:
\begin{equation}
	\mathcal{F}_{ij}^{\rm RSD}=\frac{V_{\text {s}}}{8 \pi^{2}} \int_{-1}^{+1} d \mu \int_{k_{\min }}^{k_{\max }} k^{2} d k F_{ij}^\text{RSD}(k,\mu)
\end{equation}
where the Fisher Matrix \citep{2016MNRAS.457.2377Z}
\begin{equation}
	F_{ij}^\text{RSD}(k,\mu)=\left(\frac{\bar{n}_\text{g}P^\text{RSD}_\text{gg}}{\bar{n}_\text{g}P_\text{gg}^\text{RSD}+1}\right)^2\frac{\partial\ln P_\text{gg}^\text{RSD}}{\partial p_i}\frac{\partial\ln P_\text{gg}^\text{RSD}}{\partial p_j}
\end{equation}

We then conduct a Fisher analysis around the fiducial parameters \(\{ b_\text{v} = 1, f = 0.53, \sigma_8 = 0.81 \}\) at \(z = 0\). For the galaxy bias, we employ the model \(b_\text{g} = 1 + 0.84z_\text{g}\). { For a CSST-like survey, we assume an optimistic case with an effective growth rate \( f^0_\text{eff} = 0.7 \) and a system density \( N_\text{sys} = 10^4 \, \text{(h/Mpc)}^3 \). }

The resulting uncertainties for \(f\sigma_8\) at three different redshifts are illustrated in Figure \ref{fig:fs8}. As anticipated, the RSD measurement provides highly accurate constraints on the growth rate, with uncertainties around \(\sim 10^{-2}\) (gray dashed lines). Higher redshifts of the spectroscopic galaxies result in even better performance, further reducing the uncertainty. { In contrast, incorporating the additional prior \(b_\text{v}\) from FRB DMs into kSZ measurements does not significantly constrain \(f\sigma_8\) (green line) when the number of FRBs is small. Therefore, when combining RSD and kSZ measurements (red line), the constraint on \(f\sigma_8\) shows only a slight improvement. However, the constraint is significantly enhanced when a large number of FRBs are available and \(\sigma_{\rm DM}\) has a small dispersion. When \(N_{\rm FRB} > 10^5\) and \(\sigma_{\rm DM} = 100 \, \text{pc/cm}^3\), the uncertainty in kSZ measurements exceeds that of RSD measurements. Ultimately, with the support of FRB DMs, the combined precision of growth rate measurements from RSD and kSZ can reach 0.1\%. Throughout all figures in this paper, kSZ refers to the combined measurement of kSZ and FRBs. The true kSZ-only scenario is not shown due to optical depth degeneracy, which introduces significant errors in measuring the cosmic growth rate using the kSZ effect.}


\section{Constraints from Growth Rate}\label{sec:constraint}

In this section, utilizing the measurements of \(f\sigma_8\) at three redshifts, we conduct a global analysis to constrain the cosmological parameters within three dark energy models: the \(\Lambda\)CDM model, the constant equation of state (EOS) model, the dynamical model, and modified gravity model. For the reference case, we select the FRB parameters \(N_{\rm FRB} = 10^6\) and \(\sigma_{\rm DM} = 300 \, \text{pc/cm}^3\). The resulting uncertainties in \(f\sigma_8\) at three redshifts from the RSD-only and combined RSD+kSZ measurements are listed in Table \ref{tab:fs8_fid}.

\subsection{MCMC baseline}\label{sec:baseline}

\begin{table}
\caption{The uncertainties of $f\sigma_8$ at three redshifts for the MCMC analysis from the RSD-only and combined RSD+kSZ measurements in the reference case $N_\text{FRB}=10^6$ and $\sigma_\text{DM}=300 \text{pc}/\text{cm}^3$.}
\label{tab:fs8_fid}
\begin{tabular}{cccc}
\hline
\multicolumn{1}{r}{} & $\sigma(f\sigma_8)_R$ & $\sigma(f\sigma_8)_{R+k}$ \\ \hline
$z_1$=0.15             & 0.03762   & 0.00486    \\
$z_2$=0.45             & 0.01312   & 0.00298   \\
$z_3$=0.75             & 0.00801   & 0.00289    \\ \hline
\end{tabular}
\end{table}

Since the cosmic growth rate depends on the fiducial model used by collaborations to convert redshifts to distances, we need to correct for the Alcock-Paczynski (AP) effect \citep{2019ApJ...887..125L}. This correction is made by rescaling the growth-rate measurements by the ratio of \(H(z)d_A(z)\) for the model in question to that of the fiducial cosmology. We define this ratio as:
\begin{equation}
\text{Ratio}(z) = \frac{H(z)d_A(z)}{H^{\text{fid}}(z)d^{\text{fid}}_A(z)},
\end{equation}
where \(H(z)\) is the Hubble parameter, and \(d_A(z)\) is the angular diameter distance. Consequently, we incorporate the following \(\chi^2\) function, as proposed by \cite{2017PhRvD..96b3542N}, into the \texttt{cobaya} package \citep{2021JCAP...05..057T} for the Markov Chain Monte Carlo (MCMC) analysis:
\begin{equation}\label{eq:lik}
 \chi^2 = \sum_i \frac{(f\sigma_8(z_i))_\text{Data} - \text{Ratio}(z_i)(f\sigma_8(z_i))_\text{Th}}{\sigma_i^2(f\sigma_8)},
\end{equation}
where \((f\sigma_8(z_i))_\text{Data}\) represents the observed values, and \((f\sigma_8(z_i))_\text{Th}\) represents the theoretical predictions. To compute the theoretical values of \(f\sigma_8\) at different redshifts, we solve the differential equation for the growth factor \(D(a)\) and calculate \(f(z)\) and \(\sigma_8(z)\) using the relations \(f = d\ln{D}/d\ln{a}\) and \(\sigma_8(a) = D(a)\sigma_8(a=1)\).

\begin{figure}[t]
    \centering
    \includegraphics[width=3in]{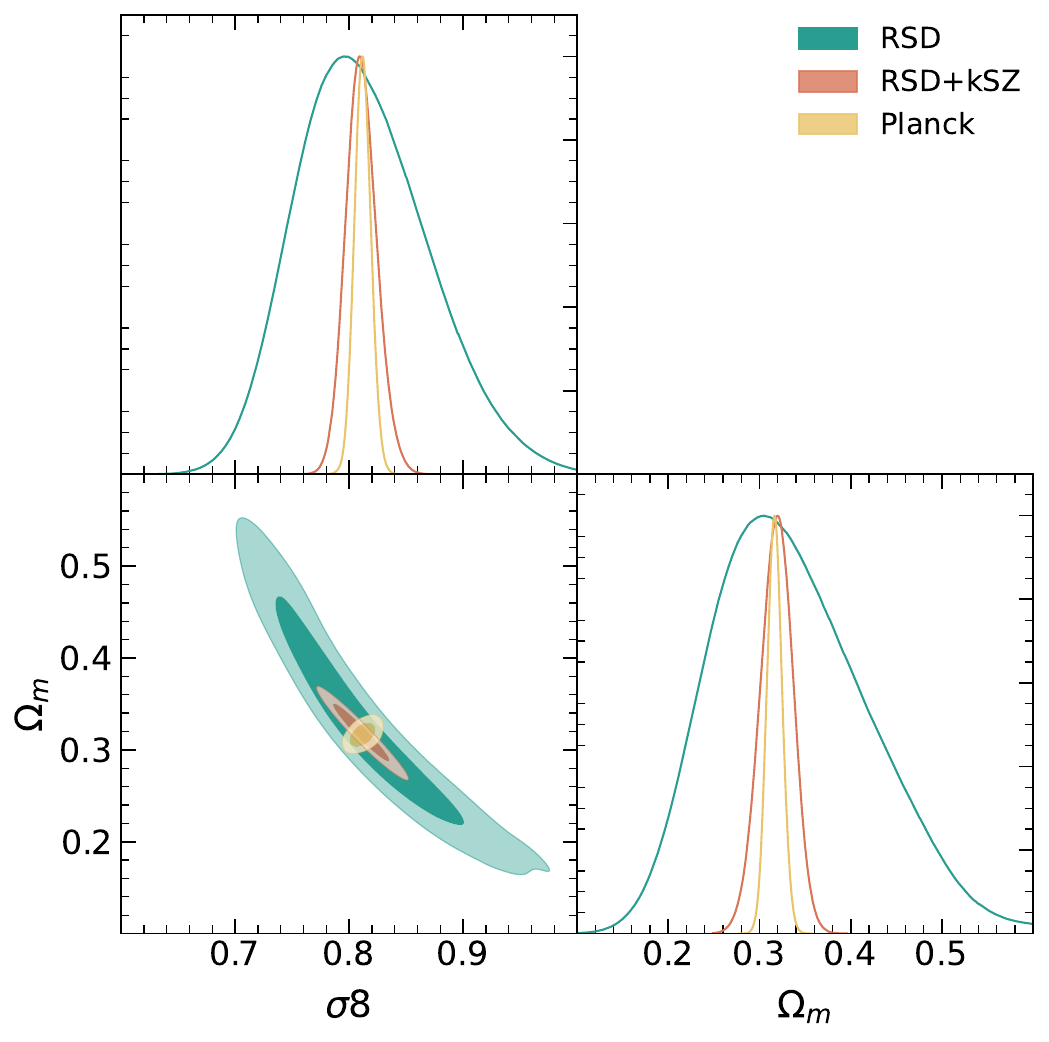}
    \caption{ The marginalized posterior contour maps and 1D probability density functions for \(\sigma_8\) and \(\Omega_\text{m}\) in the \(\Lambda\)CDM model. They are derived using three different datasets: RSD-only, combined RSD+kSZ, and the Planck measurements.}
    \label{fig:mcmc1}
\end{figure}

\subsection{$\Lambda$CDM model}\label{sec:results}

Firstly, we consider the simplest $\Lambda$CDM model and investigate the constraining power of the growth rate from the RSD and kSZ measurements in detail.

In Figure \ref{fig:mcmc1}, we present the 1D and 2D constraints on \(\Omega_\text{m}\) and \(\sigma_8\) derived from the growth rate data obtained via RSD and kSZ measurements, alongside constraints from Planck observations. The results indicate that when using only the RSD measurement data, the constraining power on these parameters is limited. However, incorporating kSZ information, particularly with the additional data from FRBs, significantly enhances the constraints, reflecting the substantial impact of the large number of FRBs on the analysis. For comparison, we also include constraints from the Planck 2018 temperature and polarization data \citep{aghanim2020planck}. The comparison shows that the combined RSD+kSZ measurements, when supplemented with FRB DMs data to obtain a more precise constraint on \(P_\text{ge}\), provide constraints on \(\Omega_\text{m}\) and \(\sigma_8\) that are comparable in strength to those derived from the Planck data.

\begin{figure}[t]
    \centering
    \includegraphics[width=3in]{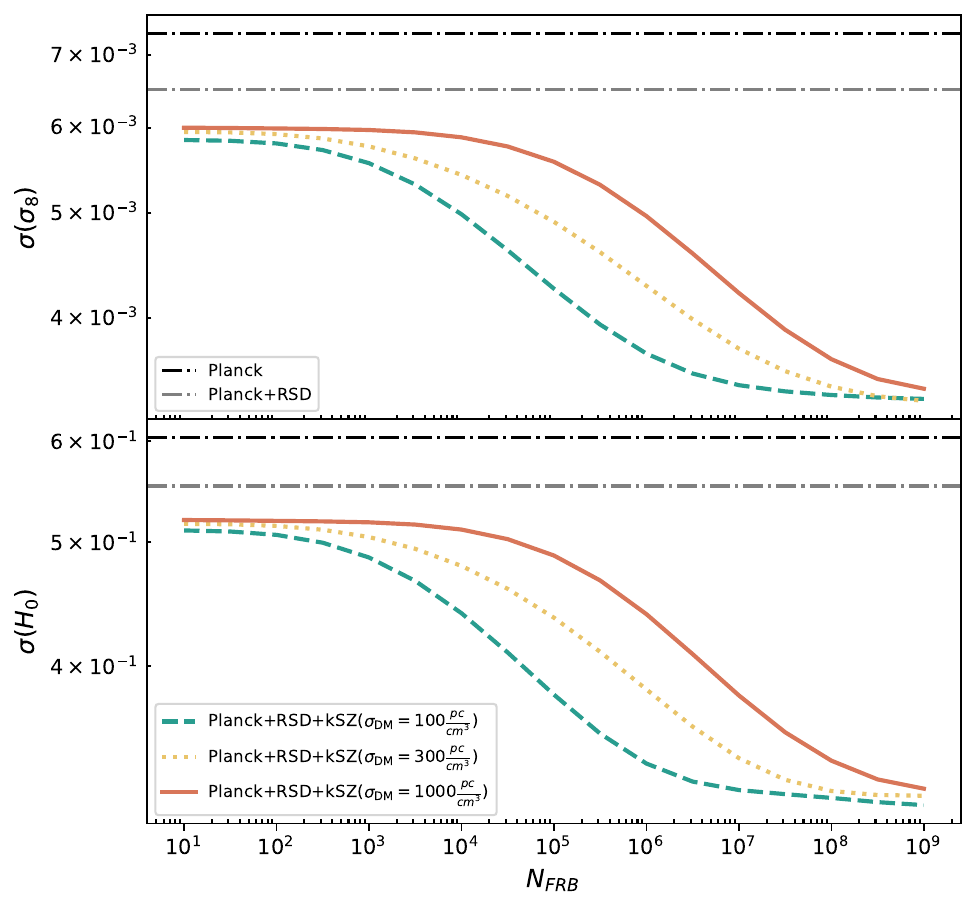}
    \caption{The uncertainties of $\sigma_8$ and $H_0$ in the $\Lambda$CDM model from different data combinations. The blank and grey dash-dotted lines represent the limitations from the Planck-only and Planck+RSD datasets, respectively. We also show the constraints from the Planck, RSD and kSZ data together.}
    \label{fig:mcmc2}
\end{figure}


Furthermore, we integrate the Fisher Matrix information from the growth rate of the combined RSD+kSZ measurements with the Planck data to jointly constrain the cosmological parameters \(\sigma_8\) and \(H_0\), as illustrated in Figure \ref{fig:mcmc2}. Using the Planck data alone, the standard deviations of \(\sigma_8\) and \(H_0\) are 0.007 and 0.6, respectively. Including the growth rate measurement from the RSD effect slightly improves these limits. When we incorporate the growth rate measurement from the kSZ effect, the constraints on \(\sigma_8\) and \(H_0\) are significantly enhanced, particularly when the number of FRBs exceeds \(10^5\). Ultimately, the combined accuracy of \(\sigma_8\) and \(H_0\) from the growth rate measurements of RSD and kSZ, along with the Planck data and supported by a large number of FRB DMs, can be improved by a factor of 2 compared to those derived from the Planck data alone.

\begin{figure*}[t]
    \centering
    \includegraphics[width=3.3in]{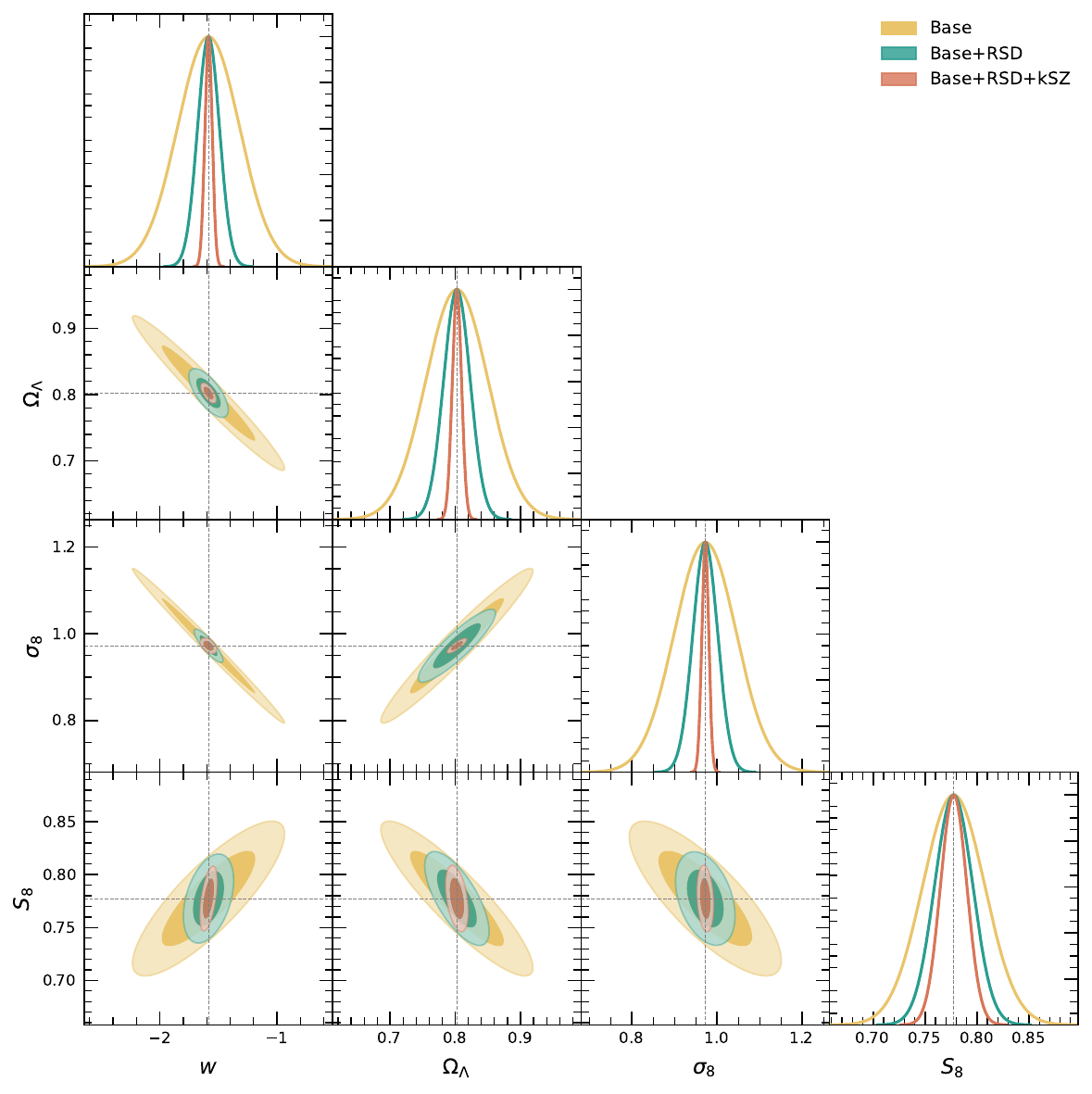}
    \includegraphics[width=3.3in]{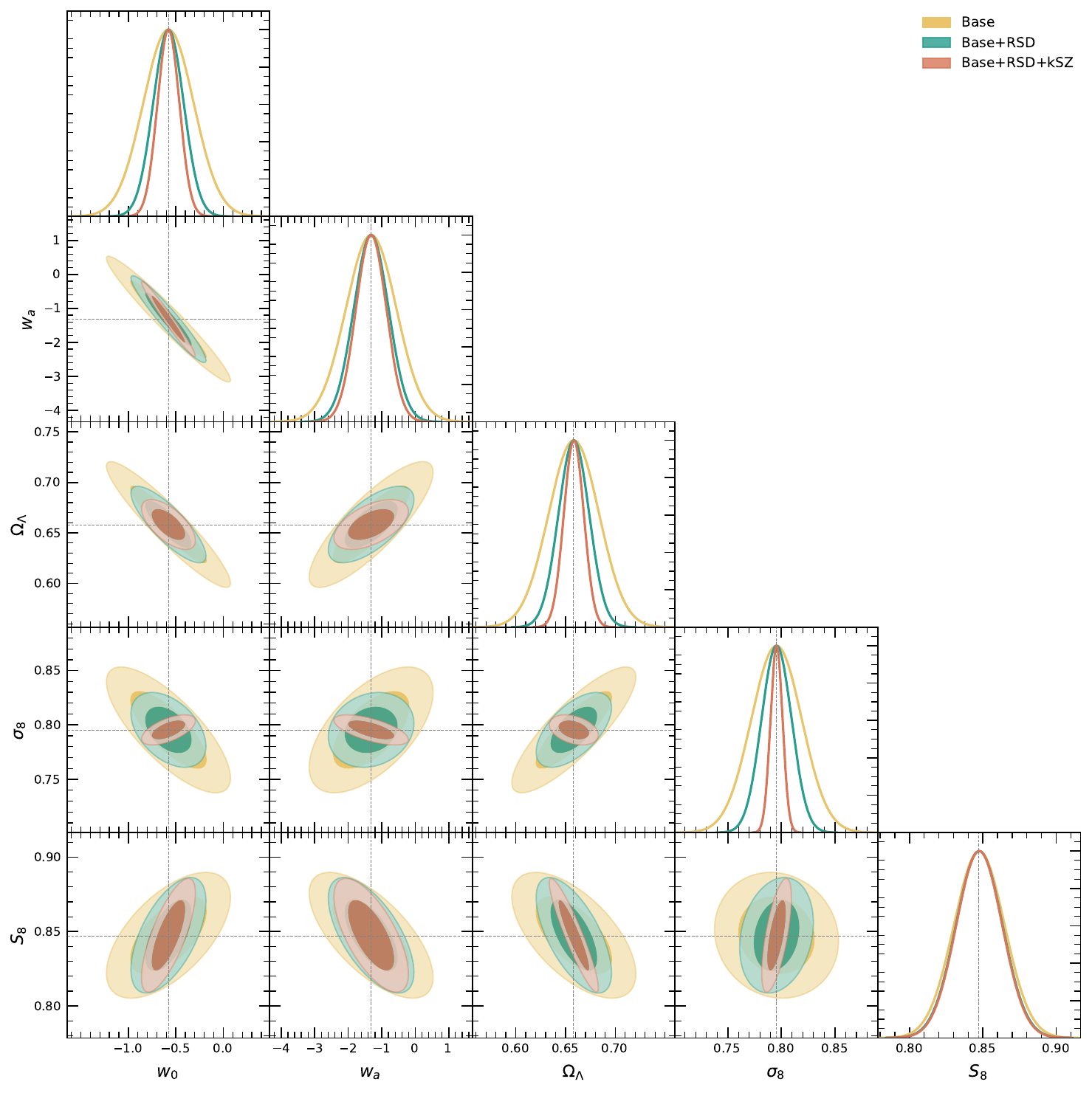}
    \caption{The marginalized posterior contour maps and 1D probability density functions for related cosmological parameters in the wCDM (left panel) and CPL (right panel) models. They are derived using three different datasets: Base only, Base+RSD and Base+RSD+kSZ.}
    \label{fig:wcdm}
\end{figure*}

The tensions in cosmological parameters, particularly the Hubble constant \citep{2023ARNPS..73..153K, 2019PhRvL.122v1301P} and the amplitude of matter density fluctuations on an 8 Mpc/h scale \citep{2023A&ARv..31....2H, PhysRevD.107.123538} are among the most intriguing challenges in contemporary cosmology. For the Hubble constant tension, the Planck satellite \citep{2020A&A...641A...6P} estimates the Hubble constant to be \(67.4\pm0.5 \, \text{km/s/Mpc}\) assuming the standard $\Lambda$CDM model. On the other hand, local measurements using the Tip of the Red Giant Branch (TRGB) method and Type Ia supernovae (SNe Ia) yield a significantly higher value of \(72.4\pm2.0 \, \text{km/s/Mpc}\) \citep{2019ApJ...886...61Y}, which have a discrepancy of about $\sim 5\sigma$ with the Planck result. For the $\sigma_8$ tension, the Planck data \citep{2020A&A...641A...6P} also provides an estimate for $\sigma_8=0.811\pm0.007$ at $z=0$. In contrast, the Dark Energy Survey (DES) Year 3 (Y3) data \citep{2022PhRvD.105b3520A}, which uses large-scale structure and weak lensing observations, suggests a lower value of $\sigma_8=0.733^{+0.039}_{-0.049}$. This more than $2\sigma$ tension with Planck indicates that the growth of structures might be slower than what the CMB-based $\Lambda$CDM model predicts. Based on above calculations, by obtaining more precise measurements of the growth rate and other cosmological parameters, future surveys like CSST and the utilization of FRBs could help resolve or at least shed more light on these tensions. Improved constraints on $H_0$ and $\sigma_8$ will allow cosmologists to either confirm the standard $\Lambda$CDM model or necessitate new physics to explain the discrepancies. This is crucial for understanding the true nature of dark energy, dark matter, and the overall evolution of the universe.

\subsection{Beyond \(\Lambda CDM model\)}

Besides the $\Lambda$CDM model, the dynamics of dark energy is crucial for understanding its nature. Therefore, in our study, we also explore two dynamical dark energy models: the constant equation of state model (wCDM) and the dynamical model characterized by the Chevallier-Polarski-Linder (CPL) parametrization, \( w(a) = w_0 + w_a(1-a) \) \citep{2001IJMPD..10..213C,2003PhRvL..90i1301L}.

We start with the wCDM model. In the left panel of Figure \(\ref{fig:wcdm}\), we present the 1D and 2D constraints on the parameters \(w\) and \(\sigma_8\) for the wCDM model using various data combinations. We choose the Planck data as the baseline. The Planck data alone exhibits a strong degeneracy between \(w\) and \(\sigma_8\), resulting in less precise constraints within the wCDM framework. However, when we incorporate growth rate information from the RSD effect, this degeneracy is significantly reduced, owing to the direct information on \(\sigma_8\) provided by the growth rate measurements. Moreover, when we include the growth rate data from the kSZ effect, enhanced by FRB measurements with \(N_{\rm FRB} = 10^6\) and \(\sigma_{\rm DM} = 300 \, \text{pc/cm}^3\), the improvement is even more pronounced and the degeneracy between \(w\) and \(\sigma_8\) is nearly eliminated. As a result, the constraint on \(w\) is dramatically improved, with the uncertainty reduced by a factor of 10. This demonstrates that FRB measurements, by offering additional and independent constraints on the distribution of matter and the growth of cosmic structures, lead to significantly tighter constraints on the nature of dark energy within the wCDM model.

Finally, we conduct similar analyses within the dark energy model using the CPL parametrization. Since the Planck data alone are insufficient to provide reasonable constraints on the CPL parameters, we use the combination of Planck and BAO data as the baseline for the subsequent discussions. Given that BAO measurements already encompass extensive information about large-scale structures, we do not anticipate significant improvements in parameter constraints from the growth rate measurements of RSD and kSZ. However, when these growth rate measurements are incorporated, particularly with a large number of FRBs, the degeneracy between dark energy parameters and \(\sigma_8\) can be effectively resolved. As a result, the constraints on dark energy parameters are notably enhanced, especially the parameters $w_0$ and $w_a$ whose uncertainties could be reduced by a factor of $\sim 2$. 

The enhancement of constraints on dark energy parameters, especially for \(w_0\) and \(w_a\), can provide significant insights into the dynamical nature of dark energy. By reducing the uncertainties in these parameters, it becomes possible to explore the evolution of the equation of state of dark energy more precisely. Reducing the uncertainties can also help verify whether dark energy exhibits a crossing behavior \citep{2005PhLB..607...35F, 2010PhR...493....1C}, where the equation of state transitions from \(w > -1\) (quintessence-like) to \(w < -1\) (phantom-like) over time. This crossing behavior is a critical feature in understanding the underlying physics of dark energy and its role in the accelerated expansion of the universe. By leveraging future large-scale surveys and FRB measurements, it is expected that these enhanced constraints will allow for a deeper investigation into whether dark energy remains constant or evolves dynamically, potentially leading to breakthroughs in cosmology  \citep{2023PhRvD.108j3519W, 2020PhR...857....1F}.

\subsection{Modified gravity}

In addition to dark energy, modified gravity is considered another potential mechanism for explaining the late-time accelerated expansion of the Universe. However, in this analysis, we adopt the \(\Lambda\)CDM background to examine the effects of modified gravity. In the conformal Newtonian gauge, many modified gravity models characterize the perturbative scalar gravitational potential \(\Phi\) and the curvature perturbation \(\Psi\). We follow the \(\mu - \eta\) parametrization as formulated in Planck collaboration studies \citep{2020A&A...641A...6P, ade2016planck}:
\begin{equation}
\begin{aligned}
    & k^{2} \Psi = -\mu(a, k) 4 \pi G_{\mathrm{N}} a^{2} [\rho \Delta + 3(\rho + P) \sigma], \\
    & k^{2} [\Phi - \eta(a, k) \Psi] = \mu(a, k) 12 \pi G_{\mathrm{N}} a^{2} (\rho + P) \sigma,
\end{aligned}
\end{equation}
where \(\mu(a,k) = G_{\mathrm{eff}}(a,k)/G_{\mathrm{N}}\) modifies the Poisson equation for \(\Psi\), and \(\eta(a,k)\) accounts for an effective additional anisotropic stress. When \(\mu = \eta = 1\), General Relativity (GR) is recovered. Both \(\mu\) and \(\eta\) are \(k\)-dependent, and as shown in \cite{ade2016planck}, introducing scale dependence in \(\mu\) and \(\eta\) does not significantly reduced \(\chi^2\) compared to the scale-independent scenario. Therefore, we focus on deviations from standard GR by adopting a late-time parametrization for \(\mu\) and \(\eta\), as described in \cite{casas2017linear}:
\begin{equation}
\begin{aligned}
    \mu(a) &\equiv 1 + E_{11} \Omega_\text{de}(a), \\
    \eta(a) &\equiv 1 + E_{22} \Omega_\text{de}(a),
\end{aligned}
\end{equation}
where \(\Omega_\text{de}(a)\) is the dark energy density, and \(E_{11}\) and \(E_{22}\) are parameters that describe the proportional effect of modified gravity on clustering and anisotropic pressure, respectively, in relation to \(\Omega_\text{de}(a)\).

An essential factor in constraining modified gravity is the growth rate of perturbations, typically characterized by \(f\sigma_8\), since modifications to GR are anticipated to affect this growth rate \citep{mueller2018clustering}. This effect is described by the following equation, which holds in many modified gravity theories \citep{2019PhRvD.100j3532M}:
\begin{equation}
\delta^{\prime\prime}(a) + \left(\frac{3}{a} + \frac{H^{\prime}(a)}{H(a)}\right) \delta^{\prime}(a) - \frac{3}{2} \frac{\Omega_{\mathrm{m}} G_{\mathrm{eff}}(a,k)/G_{\mathrm{N}}}{a^5 H(a)^2 / H_0^2} \delta(a) = 0,
\end{equation}
where \(\delta(a)\) is the growth factor, \(G_{\mathrm{N}}\) is the Newtonian constant, and \(G_{\mathrm{eff}}\) is the effective Newton constant representing the influence of modified gravity.

Therefore, the \(f\sigma_8\) results presented in this study can be integrated with the Planck 2018 temperature, polarization, and lensing data \citep{aghanim2020planck} to jointly constrain the modified gravity parameters \(\mu\) and \(\eta\). The likelihood for \(f\sigma_8\) follows Eq.\ref{eq:lik}, and the associated uncertainties are obtained from Table \ref{tab:fs8_fid}. This study adopts the \(\Lambda\)CDM background and the \(\mu-\eta\) parametrization for the modified gravity model. MCMC simulations were conducted using \texttt{MGCobaya} \citep{wang2023new, zucca2019mgcamb, hojjati2011testing, zhao2009searching}. Constraints were first derived using Planck data, and post-processing was subsequently performed by integrating \(f\sigma_8\) data.

\begin{figure}
    \centering
    \includegraphics[width=3in]{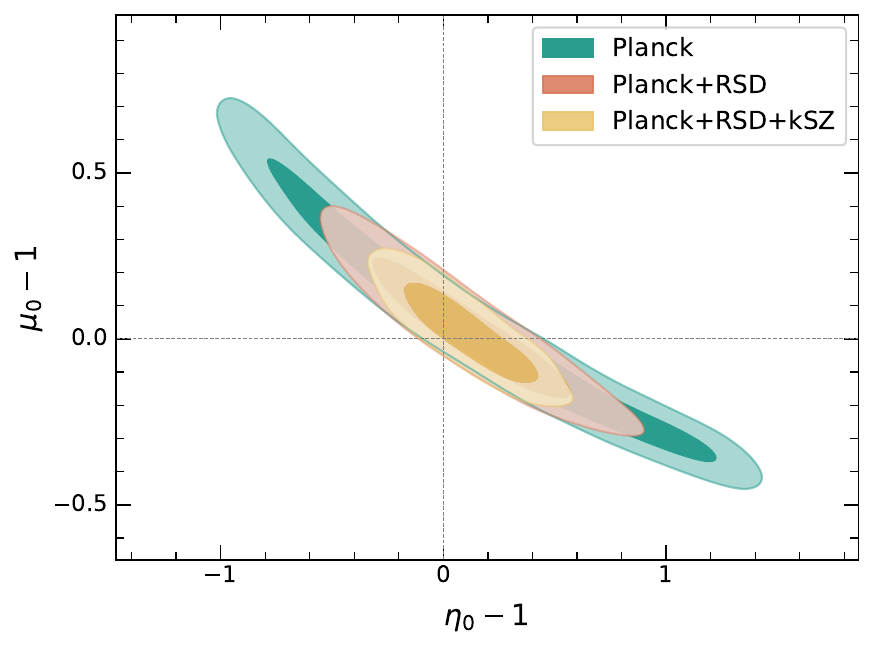}
    \caption{The marginalized posterior contour map for $\mu_0-1$ and $\eta_0-1$ in the $\Lambda$CDM background and $\mu-\eta$ parameterization. They are derived using three different datasets: Planck-only (TTTEEE+lowE+lensing), Planck+RSD and Planck+RSD+kSZ.}
    \label{fig:mg}
\end{figure}

Figure \ref{fig:mg} displays the 2D constraints for \(\mu_0 - 1\) and \(\eta_0 - 1\) under the \(\Lambda\)CDM background using the \(\mu-\eta\) parametrization, where the subscript \(0\) denotes the present time. The contours are obtained from three datasets: Planck-only, Planck + RSD, and Planck + RSD + kSZ. The figure demonstrates that, compared to the Planck-only results, the constraints progressively tighten with the addition of supplementary data. With Planck data alone, the 1-\(\sigma\) errors for \(\mu_0 - 1\) and \(\eta_0 - 1\) are 0.27 and 0.6, respectively. The addition of \(f\sigma_8\) data from RSD measurements enhances the constraining power by a factor of two. When kSZ measurements incorporating FRB information are considered, substantial improvements are achieved in constraining the modified gravity parameters for \(N_\text{FRB} = 10^6\) and \(\sigma_\text{DM} = 300 \, \text{pc}/\text{cm}^3\), resulting in standard deviations of 0.10 and 0.19 for \(\mu_0 - 1\) and \(\eta_0 - 1\), respectively.

In fact, the constraining power of FRB is comparable with that of DESI data. \cite{ishak2024modified} used DESI (FS + BAO), CMB (PR3), DESY3, and DESY5 SN data to obtain \(\mu_0 - 1 = 0.020^{+0.19}_{-0.24}\) and \(\eta_0 - 1 = 0.09^{+0.36}_{-0.60}\). More stringent constraints on modified gravity parameters such as \(\mu\) and \(\eta\) can help elucidate their contributions to deviations from the standard cosmological model, improving our understanding of dark energy and modified gravity theories, and providing insights into the nature of cosmic acceleration \citep{ade2016planck}.




\section{Conclusion and discussion}\label{sec:discussion}

The growth rate measurements obtained from traditional methods, such as the kSZ effects, are significantly influenced by the large uncertainty of the small-scale galaxy-electron power spectrum \(P_{\text{ge}}(k_s)\). Improving our understanding of \(P_{\text{ge}}(k_s)\) is crucial for more accurate growth rate determinations. By correlating the dispersion measure (DM) of FRBs with galaxy surveys, it is possible to obtain precise measurements of the electron-galaxy cross-power spectrum \(P_{\text{ge}}(k_s)\) across a wide range of scales. This can break the degeneracy between \(P_{\text{ge}}(k_s)\) and the galaxy-velocity power spectrum \(P_{\text{gv}}\), which often limits traditional methods. Thus, FRBs represent a powerful tool for advancing our understanding of the universe's large-scale structure and growth.

In this paper, we employ the kSZ and RSD effects, combined with the Fast Radio Bursts (FRBs) dispersion measure (DM) power spectrum, to achieve high-precision measurements of the product of the cosmic growth rate \(f\) and \(\sigma_8\). The unknown velocity biases (\(b_\text{v}\)) in traditional methods, such as kSZ effect, can introduce uncertainties into the growth rate measurements, particularly when relying solely on galaxy-CMB cross-correlations to reconstruct the small-scale electron-galaxy power spectrum \(P_{\text{ge}}(k_s)\). We highlight that FRBs provide independent and direct measurements of electron density along their lines of sight allows for a more precise determination of \(P_{\text{ge}}(k_s)\). This improved precision over the traditional galaxy-CMB approach reduces the uncertainties in the reconstructed power spectrum, making the cosmological application of the kSZ effect much more feasible. 

By incorporating the information from FRBs into traditional kSZ and RSD methods, we can significantly enhance the precision of growth rate predictions, as demonstrated through Fisher forecast analyses. The resulting precise growth rate measurements will play a crucial role in constraining cosmological models, especially in addressing existing tensions in cosmological parameters such as \(\sigma_8\). Thus, FRBs represent a transformative tool for improving the accuracy of large-scale structure growth rate measurements.

Here we summarize our main conclusions in more detail:

\begin{enumerate}
	\item We calculated the 1$\sigma$ uncertainty in the growth rate by combining galaxy surveys with the FRBs dispersion measure (DM) field. Our findings indicate that incorporating FRBs provides valuable additional information, which significantly reduces the uncertainty in the measurement of \(f\sigma_8\). Notably, the combination of RSD and kSZ measurements benefits greatly from the presence of FRBs, leading to a marked decrease in uncertainty. The reduction is especially pronounced with a smaller dispersion measure \(\sigma_\text{DM}\) and a larger number of FRBs. For instance, when the number of FRBs \(N_\text{FRB}\) exceeds \(10^5\) and \(\sigma_\text{DM}\) is 100 \(\text{pc}/\text{cm}^3\), the kSZ effect's contribution to the \(f\sigma_8\) measurement becomes more substantial, thereby providing tighter constraints. The most optimistic scenario suggests that with these conditions, the uncertainty in \(f\sigma_8\) could be reduced to the order of 0.1\%. This significant improvement in precision underscores the potential of combining FRBs with traditional galaxy surveys and kSZ measurements for more accurate assessments of cosmic structure growth rates and better constraints on cosmological parameters.

	\item For the \(\Lambda\)CDM model, we calculated the 1$\sigma$ uncertainties for the parameters \(\sigma_8\) and \(H_0\) using the derived growth rate from both traditional methods and the combined approach involving FRBs. We observed that the growth rate measurements using RSD show a slight increase compared to the Planck results. However, the inclusion of kSZ measurements, especially with \(N_\text{FRB} = 10^5\), leads to a significant enhancement in the constraints on these parameters. Specifically, the precision on \(H_0\) and \(\sigma_8\) improves by a factor of approximately two when incorporating kSZ data with FRBs. This improvement indicates that future surveys, such as the CSST and larger numbers of FRBs, have the potential to resolve the current tensions in \(H_0\) and \(\sigma_8\). These tensions, which currently stand at approximately 5$\sigma$ for \(H_0\) and 2$\sigma$ for \(\sigma_8\), could be significantly mitigated by leveraging the enhanced precision provided by these advanced observational techniques.

	\item For the wCDM and CPL parametrizations, we concentrated on constraining the dark energy parameters and \(\sigma_8\). In the wCDM framework, with \(N_\text{FRB} = 10^6\) and \(\sigma_\text{DM} = 300 \text{pc}/\text{cm}^3\), our results show an improvement by about a factor of 10 over the constraints on \(\omega\) provided by Planck, with the degeneracy between \(\omega\) and \(\sigma_8\) being nearly resolved. For the CPL parametrization, we observe a twofold improvement in the constraints on both \(w_0\) and \(w_a\) compared to the Planck+BAO data combination. Increasing the number of FRBs will further enhance our ability to study the nature and evolution of dark energy, providing deeper insights into dark energy models.

	\item The \(\mu-\eta\) parametrization was employed to model modified gravity, with the parameters \(\mu_0 - 1\) and \(\eta_0 - 1\) being constrained using \(N_\text{FRB} = 10^6\) and \(\sigma_\text{DM} = 300 \, \text{pc}/\text{cm}^3\). By combining \(f\sigma_8\) data from RSD and kSZ measurements with Planck data, the resulting constraints on the modified gravity parameters achieved standard deviations of 0.10 and 0.19, respectively. This represents approximately a twofold improvement over the constraints derived from Planck data alone. These findings highlight that more extensive and higher-precision FRB observations can significantly enhance the accuracy of constraints on modified gravity models, offering a deeper understanding of their impact on cosmic acceleration.
 
\end{enumerate}

{ However, it is important to note that the number of FRBs considered in this study far exceeds the current observational tally. { Nevertheless, significant progress has been made in FRB detection. The number of observed FRBs has increased dramatically, primarily due to the CHIME radio telescope \citep{2023ApJ...947...83C}. Future observations with SKA-mid are expected to play a crucial role, with its target event rate inferred from existing FRB surveys such as SKA1-mid \citep{Zhang:2023gye}. Based on Parkes survey data, the all-sky FRB event rate during the SKA-mid phase is projected to reach approximately \(10^6\) events per sky per day. Given the SKA observation plan—where FRBs are monitored for one-fifth of the year over a \(20\,\text{deg}^2\) survey area—around \(10^5\) FRBs are expected to be detected. Consequently, a total of \(10^6\) FRBs could be observed over a decade.} 

Additionally, ongoing surveys such as DAS \citep{2019BAAS...51g.255H} and FAST \citep{2011IJMPD..20..989N} continue to expand the FRB dataset, making a sample size of \(10^6\) increasingly feasible. Moreover, advancements in machine learning \citep{2024ApJ...960..128K} are enhancing FRB signal detection, while improvements in dispersion measure (DM) precision and a better understanding of the redshift-DM relation \citep{2024ApJ...965...57B} will further refine the results of this study.}

The high sensitivity of the DM to the free electron density along the line of sight, combined with the role of FRBs as backlights, offers valuable insights into components that are otherwise challenging to detect. This capability significantly broadens their applications in cosmology without necessitating extensive additional simulations. As the detection of FRBs continues to increase, there will be opportunities to explore fundamental physics, such as the fine-structure constant \citep{2024arXiv240611691L}, and to place more precise constraints on cosmological parameters, including the Hubble constant \citep{2022MNRAS.516.4862J}, cosmic baryon distribution \citep{2024arXiv240200505K}, and baryonic feedback models \citep{2024arXiv240308611T}.

\begin{acknowledgements}
This work is supported by the National Science Foundation of China under grant Nos. 12473004 and 12021003, the National Key R\&D Program of China No. 2020YFC2201603, the China Manned Space Program through its Space Application System, and the Fundamental Research Funds for the Central Universities.
\end{acknowledgements}

\bibliographystyle{spphys}       
\bibliography{template-epjc}   

%
%

\end{document}